\documentclass[epj,final]{svjour}
\usepackage{graphicx}
\usepackage{multirow}

\DeclareGraphicsExtensions{.jpg}

\newcommand{\citet}[2]{#2~\cite{#1}}
\newcommand{\etal}[0]{~et~al.}

\newcommand{\secref}[1]{Section~\ref{sec_#1}}
\newcommand{\figref}[1]{Figure~\ref{fig_#1}}
\newcommand{\tblref}[1]{Table~\ref{tbl_#1}}
\newcommand{\eqref}[1]{equation~(\ref{eq_#1})}

\newcommand{\argmax}[0]{\arg\!\max}

\newcommand{\lpa}[0]{\textit{LPA}}
\newcommand{\bpa}[0]{\textit{BPA}}
\newcommand{\bpal}[0]{\textit{BPA$_L$}}
\newcommand{\gmo}[0]{\textit{MO}}

\begin{document}



\title{Robust network community detection using balanced propagation}
\author{L. \v Subelj\thanks{\email{lovro.subelj@fri.uni-lj.si}} \and M. Bajec}
\institute{University of Ljubljana, Faculty of Computer and Information Science, Ljubljana, Slovenia}
\date{\today}

\abstract{Label propagation has proven to be an extremely fast method for detecting communities in large complex networks. Furthermore, due to its simplicity, it is also currently one of the most commonly adopted algorithms in the literature. Despite various subsequent advances, an important issue of the algorithm has not yet been properly addressed. Random (node) update orders within the algorithm severely hamper its robustness, and consequently also the stability of the identified community structure. We note that an update order can be seen as increasing propagation preferences from certain nodes, and propose a balanced propagation that counteracts for the introduced randomness by utilizing node balancers. We have evaluated the proposed approach on synthetic networks with planted partition, and on several real-world networks with community structure. The results confirm that balanced propagation is significantly more robust than label propagation, when the performance of community detection is even improved. Thus, balanced propagation retains high scalability and algorithmic simplicity of label propagation, but improves on its stability and performance.}


\maketitle


\section{\label{sec_intro}Introduction}
Complex real-world networks can comprise local structural modules (i.e., \textit{communities}~\cite{GN02}) that are groups of nodes densely connected within and only loosely connected with the rest of the network. Communities may play important roles in different real-world systems -- they can be related to functional modules in biochemical networks~\cite{PDFV05} or individuals with common interests in social networks~\cite{GN02}. Moreover, community structure also has a strong impact on dynamic processes taking place on such networks~\cite{ADP06} and can thus provide an important insight into not only structural organization but also functional behavior of various real-world systems.

As a consequence, analysis of network community structure has been the focus of recent endeavor in different fields of science. There has also been a substantial number of community detection algorithms proposed in the literature over the last years~\cite{CNM04,WH04,PDFV05,SJN06,RAK07,AK08,BGLL08,RB08,Liu10,RN10,SB11} (for a comprehensive survey see~\cite{For10}). Nevertheless, due to scalability issues, only a small minority of these algorithms can be applied to large real-world networks with several millions, billions of nodes, edges respectively.

A notable step towards this end was made by~\citet{RAK07}{Raghavan\etal}, who employed a simple \textit{label propagation} to reveal significant communities in large real-world networks. Communities are identified by propagating (community) labels among nodes, thus, each node is assigned the label shared by most of its neighbors. Due to very fast structural inference of label propagation, densely connected sets of nodes form a consensus on some particular label after only a few iterations~\cite{RAK07,SB11}. The algorithm thus exhibits near linear complexity, which makes it applicable on networks with millions of nodes in a matter of minutes~\cite{SB11}. The basic algorithm was further analyzed and refined by various authors~\cite{HCZLDF08,TK08,BC09,LHLC09,LM09b,LM09c,PCW09,PSSL09,Gre10,LM10,SB11,SB10a,YWGW10}, when, due to its simplicity, label propagation is also currently one of the most commonly adopted algorithms in the literature.

Despite the above efforts, an important issue of label propagation has not yet been properly addressed. To overcome convergence problems in some types of networks, \citet{RAK07}{Raghavan\etal} have proposed propagating labels among nodes (i.e., updating nodes' labels) in a random order. Although this updating strategy solves the aforementioned problem, introduction of randomness severely hampers the robustness of the algorithm, and consequently also the stability of the identified community structure. It has been noted that the algorithm reveals a large number of distinct community structures even in smaller networks~\cite{RAK07,TK08,LM09b,SB11}, when these structures are also relatively different among themselves~\cite{TK08,SB11}. Still, the robustness of the algorithm can also be related to the significance of community structure in a network~\cite{SB11}.

We argue that updating the nodes in some particular order can be seen as placing higher \textit{propagation preference}~\cite{LHLC09} to the nodes that are updated at the beginning, and lower propagation preference to the nodes that are updated towards the end (and updating the nodes in a random order). The order of node updates thus governs the dynamics of the algorithm in a similar manner as  (corresponding) node propagation preferences. This observation allows us to stabilize the label propagation algorithm by utilizing node preferences to counteract (i.e., balance) the randomness introduced by random node updates. The resulting algorithm is denoted \textit{balanced propagation} and differs from label propagation merely in the introduction of \textit{node balancers}.

We have evaluated the proposed algorithm on synthetic benchmark networks with planted partition, and on various real-world networks with community structure. The results confirm that balanced propagation is significantly more robust than simple label propagation, when the performance of community detection is even improved (in most cases). We also apply the algorithm to an entire European road network, which is not considered to reveal clear community structure. Nevertheless, the algorithm accurately identifies communities that correspond to different (geographical) regions of Europe, without any serious issues with stability.

The rest of the article is organized as follows. In~\secref{lp} we formally present label propagation, and review issues and advances relevant for this research. \secref{bp} introduces balanced propagation and discusses the main rationale behind it. Empirical evaluation with discussion is given in~\secref{expr} and conclusion in~\secref{conc}.


\footnotetext[1]{In directed networks, each edge is treated as undirected, and in multi-networks, multiple edges among nodes are encoded into edge weights.}

\section{\label{sec_lp}Label propagation}
Let the network be represented by a simple undirected graph $G(N,E)$, where $N$ is the set of nodes and $E$ is the set of edges\footnotemark[1]. Furthermore, let $w_{nm}$ be the weight of the edge incident to nodes $n,m\in N$. Moreover, let $c_n$ denote the community (label) of node $n\in N$ and let $\mathcal{N}(n)$ denote the set of its neighbors.

Basic \textit{label propagation algorithm} (\lpa)~\cite{RAK07} reveals network communities by exploiting the following simple procedure. At first, each node $n\in N$ is labeled with an unique label, $c_n=l_n$. Then, at each iteration, each node adopts the label shared by most of its neighbors (considering also edge weights). Hence,
\begin{eqnarray}
c_n & = & \argmax_l\sum_{m\in\mathcal{N}^l(n)}w_{nm},
\label{eq_lpa}
\end{eqnarray}
where $\mathcal{N}^l(n)$ is the set of neighbors of $n\in N$ that share label $l$ (ties are broken uniformly at random). Due to the existence of many intra-community edges, relative to the number of inter-community edges, densely connected sets of nodes form a consensus on some particular label after a few iterations. Thus, when the algorithm converges (i.e., equilibrium is reached), disconnected sets of nodes sharing the same label are classified into the same community. Due to extremely fast structural inference of label propagation, the algorithm exhibits near linear time complexity~\cite{RAK07,SB11} (in the number of edges of the network) and can easily scale to networks with millions, or even billions, of nodes and edges ~\cite{SB11,SB10a}.

\citet{LHLC09}{Leung\etal} have first noticed that label propagation can be substantially improved by increasing \textit{propagation preference} (i.e., propagation strength) from certain nodes. The updating rule of the algorithm (i.e., \eqref{lpa}) is thus rewritten into
\begin{eqnarray}
c_n & = & \argmax_l\sum_{m\in\mathcal{N}^l(n)}p_mw_{nm},
\label{eq_ppa}
\end{eqnarray}
where $p_n$ is the preference of node $n\in N$. Adequate node preferences can alter the dynamics of label propagation, in order to guide the algorithm towards a more significant community structure~\cite{SB11}. For the analysis and comparison of different node preference strategies, and corresponding algorithms, see~\cite{LHLC09,SB11,SB10a}.

Next, we also discuss two main issues of label propagation and its advances. First, consider a bipartite network with two sets of nodes, denoted red and green nodes. Further assume that, at some point of the algorithm, all red nodes share label $l_r$, and all green nodes share label $l_g$. Due to bipartite structure, at the next iteration, all red nodes will adopt label $l_g$, and all green nodes will adopt label $l_r$. Moreover, at next iteration, all nodes will recover their initial labels, failing the algorithm to converge. It should be noted that such oscillations of labels are not limited to bipartite networks, but occur in various real-world networks that are commonly analyzed in the literature.

To ensure convergence, \citet{RAK07}{Raghavan\etal} have proposed \textit{asynchronous} updating of nodes. Hence, nodes are no longer updated all together, but sequentially, in some random order. Thus, when node's label is updated, possibly already updated labels of its neighbors are considered (in contrast to \textit{synchronous} updating, where only labels from the previous iteration are considered). Although asynchronous updating eliminates aforementioned oscillations of labels, introduction of randomness severely disturbs the robustness of the algorithm, and consequently also the stability of the identified community structure. 
The stability of label propagation presents a severe issue for the algorithm, however, it has not yet been properly addressed in the past (to the best of our knowledge).

Second, consider a network with \textit{overlapping communities}~\cite{PDFV05} and let $n\in N$ be a node that has equally strong connections with two or more such communities. As ties are broken uniformly at random (see~\eqref{lpa}), label $c_n$ would then, in general, constantly change. Furthermore, when many of such nodes exist, the algorithm would obviously never converge. Again, the issue is not limited to networks with overlapping communities.

Two possible solutions have been proposed in the literature. \citet{LHLC09}{Leung\etal} suggested including label $c_n$ into the maximal label consideration (besides merely neighbors' labels), when \citet{RAK07}{Raghavan\etal} proposed a slightly modified approach. When there are multiple maximal labels (among neighbors' labels), and one of them equals the concerned label $c_n$, the node retains its label. In contrast to the former, the latter approach considers concerned label only when there indeed exist multiple maximal labels. Although both presented approaches work well for simple label propagation (i.e., \eqref{lpa}), this is not necessarily the case for different advances of the algorithm (e.g., \eqref{ppa}). Still, for the analysis in this article we adopt the approach proposed by~\citet{RAK07}{Raghavan\etal}.

In the proceeding section we revisit both issues discussed above, and propose solutions to overcome them.


\section{\label{sec_bp}Balanced propagation}
Label propagation with asynchronous updating accesses the nodes in a random order. In particular, nodes are (re)shuffled before each iteration, in order to address convergence issues in some networks. However, as already discussed in~\secref{lp}, this incorporation of randomness severely hampers the robustness of the algorithm.

The issue can be addressed in an \textit{ad hoc} fashion by simply accessing the nodes in some predefined (deterministic) order. This would clearly stabilize the algorithm, and possibly also perform well on real-world networks. We have conducted several experiments with different update orders, based on various node statistics (i.e., degree and eigenvector centrality~\cite{Fre77,Fre79}, clustering coefficient~\cite{WS98}). Exact results are omitted, however, they indicate that, although none of these deterministic orders performs well in all networks, best order commonly corresponds to node preference strategy that also performs well. For instance, when ordering the nodes based on their degrees (decreasingly) gives good results, setting propagation preferences to the degrees of the nodes (and updating them in a random order) also performs well (and vice-versa).

Based on the above discussion we pose a hypothesis that the order of node updates within asynchronous label propagation governs algorithm's dynamics in a similar manner as the corresponding node propagation preferences. Intuitively, nodes that are updated at the end of some iteration cannot efficiently propagate their final labels onward, as (most of) their neighbors have already been updated. On the other hand, a node that is considered first can possibly propagate its label to all of its neighbors, and thus form a community. Hence, nodes updated at the beginning exhibit higher propagation strength than those that are considered towards the end.

\begin{figure}[t]
\centering
\includegraphics[width=0.50\columnwidth]{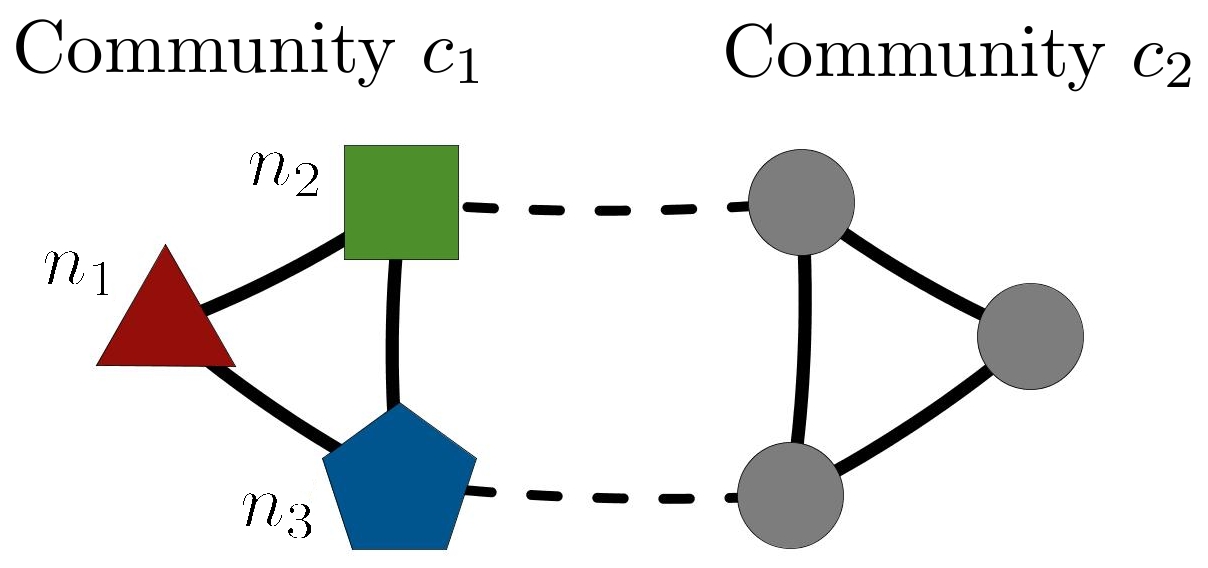}
\caption{\label{fig_toy}(Color online) Toy example network with two strong communities (inter-community edges are shown with dashed links). Node colors (shapes) indicate their community labels.}
\end{figure}

We further study the proposed hypothesis on a toy example network in~\figref{toy}. The network consists of two communities, namely $c_1$ and $c_2$, that are defined in a \textit{strong sense}~\cite{RCCLP04} (i.e., each node has more intra-community than inter-community edges). Further assume that, at some point of the algorithm, nodes in $c_1$, namely $n_1$, $n_2$ and $n_3$, are labeled with unique (community) labels, when all nodes in $c_2$ have already been classified to their right community (see~\figref{toy}).

We first analyze how different orders of node updates affect the final outcome of the algorithm. When node $n_1$ is considered first, it will adopt the label of either $n_2$ or $n_3$. Due to symmetry, we can assume that it adopts the label of node $n_2$. No matter which of the nodes $n_2$ or $n_3$ is updated next, at the end of this iteration, all nodes in community $c_1$ will be labeled with the same label (that initially belongs to node $n_2$). The outcome thus corresponds to the natural community structure of the network.

On the other hand, when node $n_1$ is updated last, the results can differ. Again, we can assume that node $n_2$ is considered before node $n_3$. If node $n_2$ adopts the label of either $n_1$ or $n_3$, the algorithm proceeds similar as above. However, node $n_2$ can also adopt the label of the second community $c_2$ (with some probability). In that case, it is straightforward to see that nodes $n_1$ and $n_3$ will also adopt the same label, thus, at the end, all nodes in the network will be classified to the same community $c_2$.

To summarize, if we first consider the core of community $c_1$ (i.e., node $n_1$), the label propagation will inevitably lead to the natural community structure of the network. However, if we access the border of community $c_1$ first (i.e., nodes $n_2$ and $n_3$), the algorithm could potentially classify all nodes into the same community (mainly due to the fact that community $c_2$ is already established). The example shows that even in such simple network, label propagation is extremely sensitive to the order of node updates.

Similar behavior as above can be observed, when we set higher propagation preference to either core or border of community $c_1$ (and update the nodes in a random order). When core node $n_1$ has the highest preference in the network, nodes $n_2$ and $n_3$ would obviously adopt the label of node $n_1$. This would unavoidably lead to identification of the natural community structure, no matter the order of updates. However, when higher preference is given to border nodes $n_2$ and $n_3$ (i.e., lowest preference is given to node $n_1$), outcome of the algorithm can again correspond to the trivial community structure, where all nodes are classified into the same community (depends on the preference of other nodes and the order of updates). We thus conclude that, at least for this toy example, order of node updates can be seen as placing higher propagation preference to the nodes that are updated first, and lower propagation preference to the nodes that are updated last.

The latter enables us to stabilize the basic label propagation algorithm. As random node updates cannot be avoided (\secref{lp}), node propagation preferences can be utilized to counteract the randomness introduced by random updates. Node preferences are thus employed to balance the algorithm (i.e., \textit{node balancers}) and are set according to the reverse order in which the nodes are assessed. This retains the dynamics of the basic algorithm, but greatly improves its robustness and the stability of the identified community structure.

Let nodes $N$ be ordered in some random way, and let $i_n$ denote the normalized position of node $n\in N$ in this order. Hence,
\begin{eqnarray}
i_n & = & \frac{\mbox{index of node }n}{|N|},
\end{eqnarray}
where $i_n\in(0,1]$. Assuming linearity, we introduce node balancers as
\begin{eqnarray}
p_n & = & i_n,
\label{eq_nb}
\end{eqnarray}
where $p_n$ is the preference of node $n\in N$ (see~\eqref{ppa}). Note that node balancers have to be recomputed at the beginning of each iteration (i.e., after each random shuffling of nodes). The resulting algorithm is else identical to the basic label propagation (with node preferences) and is denoted \textit{balanced propagation algorithm} (\bpa). Empirical evaluation in~\secref{expr} shows that balanced propagation is not only more stable than basic label propagation, but also improves its community detection. Note also that the revealed community structure could be even further stabilized by, e.g., combining multiple network partitions~\cite{SG02}.

We also analyze a variant of the algorithm, where logistic function is used to model the relation between update orders and propagation preferences (the algorithm is denoted~\bpal). Hence, node balancers are set due to
\begin{eqnarray}
p_n & = & \frac{1}{1+e^{-\beta(i_n-\alpha)}},
\label{eq_nbl}
\end{eqnarray}
where $\alpha$ and $\beta$ are parameters of the algorithm. We fix $\alpha=\frac{1}{2}$ and $\beta=5$ based on some preliminary experiments. Empirical analysis reveals that \bpal~usually performs slightly better than \bpa~(\secref{expr}).

Last, we also briefly consider the second main issue of label propagation. As already discussed in~\secref{lp}, nodes having equally strong connections with several (overlapping) communities might prevent the algorithm from converging. The problem is even enhanced in the case of balanced propagation, as random node preferences, introduced through random update orders, can extend the issue to cases, where node has only similarly strong connections with different communities. Consequently, solutions proposed in the literature~\cite{RAK07,LHLC09} do not necessarily overcome the problem in the case of balanced propagation.

Still, the true reason behind these convergence problems is the existence of overlapping communities in real-world networks. However, the purpose of this research is to address issues with random update orders, and not to extend balanced propagation to overlapping communities (see,~e.g.,~\cite{Gre10}). Thus, for the sake of the empirical analysis, we adopt the following simple approach (and limit the analysis to non-overlapping communities).

As the discussed problems of balanced propagation (i.e., \bpa~and \bpal~algorithms) are actually an artifact of node balancers, we simply discard their use, when the algorithm does not converge after at most some maximal number of iterations. Note that this is in fact identical to applying the basic label propagation (i.e., \lpa\space algorithm) afterwards, which obviously ensures the algorithm's convergence. We fix the maximum number of iterations to $100$, what should suffice for networks with almost a billion edges~\cite{SB11}.


\section{\label{sec_expr}Experiments and discussion}
First, balanced propagation was analyzed, and compared against label propagation, on synthetic benchmark networks with planted partition and on several real-world networks with community structure (sections~\ref{sec_expr_sn},~\ref{sec_expr_rwn} respectively). We address the stability of the algorithms and also the accuracy of community detection. Next, the proposed algorithm was further applied to a complete European road network, when the results are analyzed and discussed in~\secref{expr_ern}.

Due to generality, results in the following sections are assessed in terms of different measures of community structure significance. Earlier work commonly reported the \textit{modularity} $Q$~\cite{NG04} of the identified community structure. Modularity measures the significance of communities due to some \textit{null model} (which is considered to be without community structure). Commonly, a random graph with the same degree sequence is selected for the null model. Hence,
\begin{eqnarray}
Q & = & \frac{1}{2|E|}\sum_{n,m\in N}\left(A_{nm}-\frac{k_nk_m}{2|E|}\right)\delta(c_n,c_m),
\label{eq_q}
\end{eqnarray}
where $A$ is the adjacency matrix of the network, $k_n$ is degree of node $n\in N$ and $\delta$ is the Kronecker delta. Higher values represent more significant community structure ($Q\in[-1,1]$), however, recent work shows that modularity has a number of severe deficiencies~\cite{FB07,KSKK07,GMC10} and should not be considered as a reliable indicator of community structure.

For a more adequate assessment of the significance of revealed communities we also adopt the \textit{conductance} $\Phi$~\cite{Bol98}. Let $S\subset N$ be some community in the network thus $|S|\leq|N|/2$. Conductance of a set of nodes $S$ is then defined as
\begin{eqnarray}
\Phi & = & \frac{\sum_{n\in S,m\in\overline{S}}A_{nm}}{\min\{k(S),k(\overline{S})\}},
\label{eq_cond}
\end{eqnarray}
where $\overline{S}$ is the complement of $S$ and $k(S)$ is the cumulative degree of $S$ (i.e., $k(S)=\sum_{n\in S}k_{n}$). Conductance thus measures the goodness of community $S$, or equivalently, the quality of corresponding network cut $(S,\overline{S})$. Lower values represent more significant communities ($\Phi\in[0,1]$). Nevertheless, conductance cannot be easily extended to an entire community structure of a network. Thus, results are commonly assessed at different scales separately, in the form of \textit{network community profile} (\textit{NCP})~\cite{LLDM09} plots. Still, due to simplicity, we also define $\overline{\Phi}$ as the average conductance over all communities in a network.

For networks with known community structure, identified communities are also compared against the true ones. We adopt two measures from the field of information theory~\cite{Mac03}. First, \textit{normalized mutual information} (\textit{NMI})~\cite{DDDA05}, has become a \textit{de facto} standard in the community detection literature. Let $\mathcal{C}$ be a partition (i.e., communities) extracted by some algorithm, and let $\mathcal{P}$ be the known partition for some network (corresponding random variables are $C$ and $P$ respectively). \textit{NMI} of $\mathcal{C}$ and $\mathcal{P}$ is then
\begin{eqnarray}
\mathit{NMI} & = & \frac{2I(C,P)}{H(C)+H(P)},
\label{eq_nmi}
\end{eqnarray}
where $I(C,P)$ is the mutual information of the partitions (i.e., $I(C,P)=H(C)-H(C|P)$), and $H(C)$, $H(P)$ and $H(C|P)$ are standard and conditional entropies. \textit{NMI} of identical partitions equals $1$, and is $0$ for independent partitions ($\mathit{NMI}\in[0,1]$).

Second, \textit{variation of information} (\textit{VOI})~\cite{Mei07}, has several desirable properties with respect to \textit{NMI}. In particular, it is symmetric local measure that also has the properties of a distance in the space of partitions. \textit{VOI} of $\mathcal{C}$ and $\mathcal{P}$ is defined as
\begin{eqnarray}
\mathit{VOI} & = & H(C|P)+H(P|C),
\label{eq_voi}
\end{eqnarray}
thus, lower values represent better correlation between partitions. The maximum value of \textit{VOI} depends on the size of the network ($\mathit{VOI}\in[0,\log|N|]$), therefore, for meaningful comparisons, we divide the obtained values with $\log|N|$~\cite{KLN08}.


\subsection{\label{sec_expr_sn}Synthetic networks with planted partition}
\begin{figure}[t]
\centering
\includegraphics[width=1.00\columnwidth]{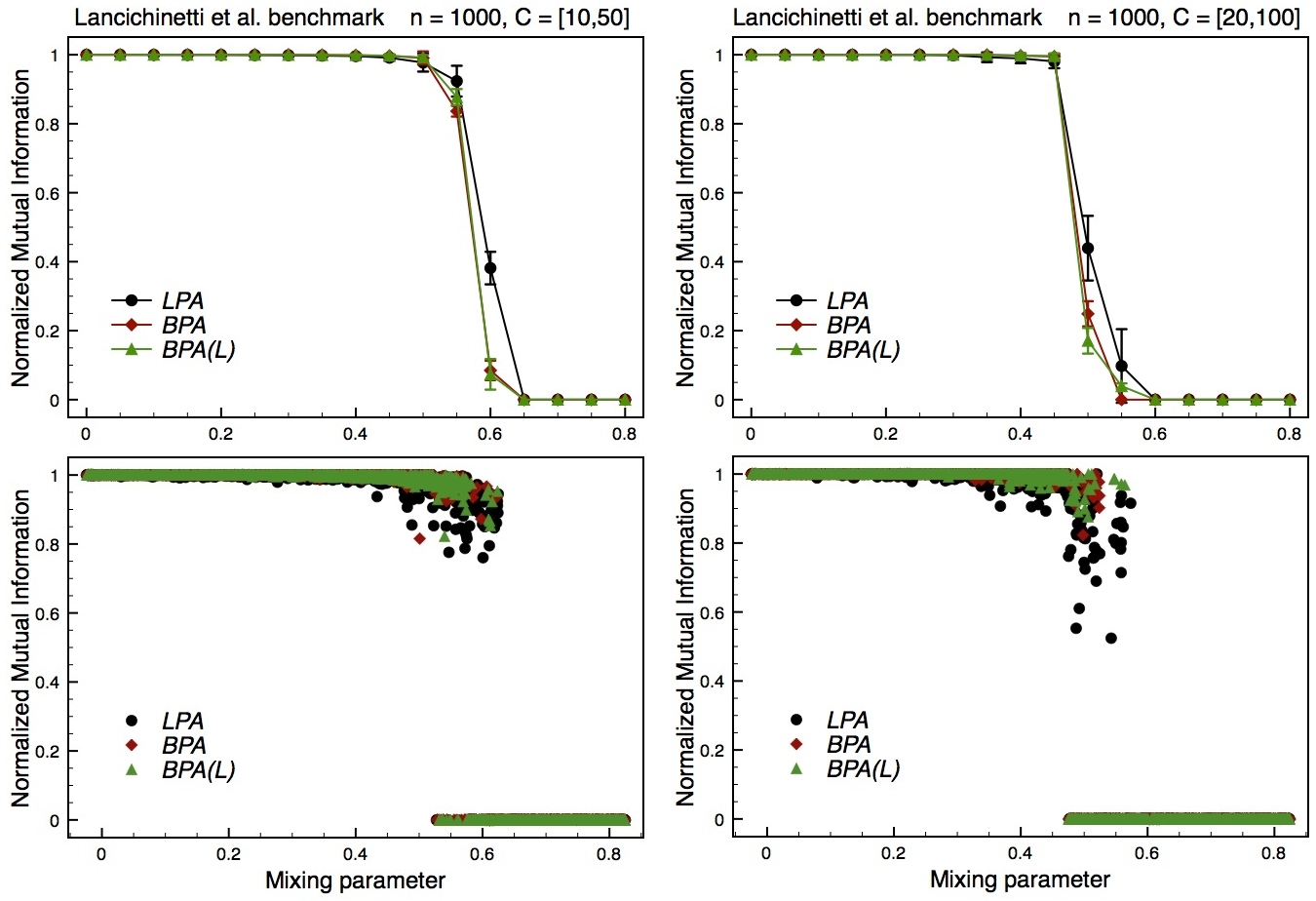}
\caption{\label{fig_bnchs}(Color online) Comparison of balanced and label propagation on synthetic benchmark networks with planted partition~\cite{LFR08}. The number of nodes is fixed to $1000$ and the sizes of communities vary between $[10,50]$ and $[20,100]$ nodes (left, right respectively). We report the averages over $100$ realizations and also the scatter plots showing individual runs (top, bottom respectively). For the former, error bars correspond to sample standard deviations computed from only nontrivial partitions (i.e., with $\mathit{NMI}>0$), and for the latter, a small amount of noise was added along the horizontal axes.}
\end{figure}

We have first analyzed the balanced propagation on a class of synthetic benchmark networks with planted partition~\cite{LFR08}. The significance of community structure is controlled by a mixing parameter $\mu\in[0,1]$, where smaller values give clearer community structure. Networks exhibit power-law degree and community size distributions, as commonly observed in real-world networks~\cite{BA99,New04}. Power-law exponents $\alpha$ are set to $2$ and $1$ respectively (i.e., $P(x)\sim x^{-\alpha}$). Moreover, we fix the number of nodes to $1000$ and vary the sizes of communities between $[10,50]$ and $[20,100]$ nodes. Results are assessed in terms of \textit{NMI} and are shown in~\figref{bnchs}.

Considering only the average performance (\figref{bnchs}, top), no clear difference between balanced propagation (i.e., \bpa~and \bpal~algorithms) and label propagation (i.e., \lpa~algorithm) is observed. However, scatter plots showing individual runs (\figref{bnchs}, bottom) reveal that there is actually a significant disparity between the approaches. When community structure is only roughly defined (i.e., for $\mu>0.5$), balanced propagation either relatively accurately identifies communities in the network (i.e., $\mathit{NMI}\approx 1$) or classifies all nodes into a single community (i.e., $\mathit{NMI}=0$). On the other hand, label propagation also commonly reports community structures, whose correspondence to the actual communities is only marginal (i.e., $\mathit{NMI}\approx 0.75$, $\mathit{NMI}\approx 0.5$ respectively). The latter is particularly apparent in the case of larger communities (note also the difference in error bars).

The results thus confirm that balanced propagation is much more robust than simple label propagation,
when the community detection strength of the basic algorithm is largely retained in the refined versions (on average). Still, to obtain results comparable with current state-of-the-art community detection algorithms (see~\cite{LF09a}), different advances of the basic approach have to be employed~\cite{SB11,SB10a}.

To further address the validity of balanced propagation, we have also applied the algorithms to a random graph \`{a}~la~Erd\"{o}s-R\'{e}nyi~\cite{ER59} that (presumably) has no community structure. The number of nodes is again fixed to $1000$, when we vary the average degree $k$ between $10$ and $100$. Both balanced propagation algorithms reveal no community structure in these networks -- all nodes are classified into a single community (or multiple communities in the case of disconnected networks) in all $100$ realizations of  random networks. On the other hand, label propagation also partitions the networks into non-trivial communities, when the average degree is small enough (i.e., for $k\leq 10$).


\subsection{\label{sec_expr_rwn}Real-world networks with community structure}
Balanced propagation was further analyzed on eight real-world networks with community structure (\tblref{rws_desc}). All these network are commonly employed in the community detection literature, and include different social, biological and technological networks. Due to simplicity, all networks were treated as unweighed and undirected.

\begin{table}[h]
\centering
\caption{\label{tbl_rws_desc}Real-world networks with community structure.}
\begin{tabular}{cccc}
\hline\noalign{\smallskip}
Network & Description & Nodes & Edges  \\
\noalign{\smallskip}\hline\noalign{\smallskip}
\textit{karate} & Zachary's karate club.~\cite{Zac77} & $34$ & $78$ \\
\textit{dolphins} & Lusseau's dolphins.~\cite{LSBHSD03} & $62$ & $159$ \\
\textit{books} & Political books.~\cite{Kre08} & $105$ & $441$ \\
\textit{football} & American football.~\cite{GN02} & $115$ & $616$ \\
\textit{jazz} & Jazz musicians.~\cite{GD03} & $198$ & $2742$ \\
\textit{elegans} & Nematode \textit{C. elegans}.~\cite{JTAOB00} & $453$ & $2025$ \\
\textit{netsci} & Network scientists.~\cite{New06a} & $1589$ & $2742$ \\
\textit{power} & U.S. power grid.~\cite{WS98} & $4941$ & $6594$ \\
\noalign{\smallskip}\hline
\end{tabular}
\end{table}

We first directly compare the stability of the revealed community structures for balanced  and label propagation (i.e., \bpa\space and \bpal, and \lpa~algorithms respectively). We apply the algorithms to each network $1000$ times and count the number of distinct community structures obtained. We also measure the pairwise \textit{VOI} of the partitions, to further evaluate the robustness of the algorithms. Due to space complexity, analysis is reduced to smaller networks (with at most hundreds of nodes). Results can be seen in~\tblref{rws_stab}.

\begin{table}[h]
\centering
\caption{\label{tbl_rws_stab}Analysis of the stability of balanced and label propagation. We report the number of distinct community structures obtained over $1000$ runs and the average pairwise \textit{VOI} of the corresponding partitions.}
\begin{tabular}{ccccccc}
\hline\noalign{\smallskip}
\multirow{2}{*}{Network} & \multicolumn{3}{c}{Distinct} & \multicolumn{3}{c}{Pairwise \textit{$VOI$}} \\
 & \lpa & \bpa & \bpal & \lpa & \bpa & \bpal \\
\noalign{\smallskip}\hline\noalign{\smallskip}
\textit{karate} & $184$ & $24$ & $\mathbf{19}$ & $0.276$ & $0.199$ & $\mathbf{0.192}$ \\
\textit{dolphins} & $525$ & $39$ & $\mathbf{36}$ & $0.256$ & $0.084$ & $\mathbf{0.079}$ \\
\textit{books} & $269$ & $37$ & $\mathbf{29}$ & $0.124$ & $\mathbf{0.100}$ & $\mathbf{0.100}$ \\
\textit{football} & $414$ & $180$ & $\mathbf{154}$ & $0.095$ & $0.093$ & $\mathbf{0.087}$ \\
\textit{jazz} & $63$ & $22$ & $\mathbf{20}$ & $0.107$ & $0.032$ & $\mathbf{0.029}$ \\
\textit{elegans} & $707$ & $\mathbf{76}$ & $\mathbf{75}$ & $0.124$ & $\mathbf{0.015}$ & $\mathbf{0.015}$ \\
\noalign{\smallskip}\hline
\end{tabular}
\end{table}

The analysis confirms earlier observations that basic label propagation is relatively unstable, even on smaller networks~\cite{RAK07,TK08,LM09b,SB11}. However, the latter does not hold for balanced propagation that reveals only a small number of distinct community structures in each network. In most cases, this number is for a scale smaller than in the case of label propagation. Moreover, the pairwise similarity between the structures is also significantly improved, when the same trend is observed if we measure similarity only among distinct structures (e.g., for \textit{elegans} network, average pairwise \textit{VOI} equals $0.1558$, $0.0430$ and $0.0424$ for \lpa, \bpa~and \bpal~algorithms respectively).

We conclude that balanced propagation is significantly more robust than label propagation, and can be, despite its randomized nature, considered as fairly stable. Note also that balanced propagation with logistic model (i.e., \bpal~algorithm) performs slightly better than the basic algorithm with a linear model (i.e., \bpa~algorithm).

Three of the networks in~\tblref{rws_desc}, namely \textit{karate}, \textit{dolphins} and \textit{football}, have known natural partitions into communities (that result from earlier studies). To analyze also the community detection strength of balanced propagation, we measure the \textit{VOI} between the natural partitions and those identified by different algorithms. The results appear in~\tblref{rws_comm}, when we also report the results for a classical modularity optimization algorithm (\gmo) proposed by~\citet{CNM04}{Clauset\etal} (for reference).

\begin{table}[h]
\centering
\caption{\label{tbl_rws_comm}Analysis of community detection strength of balanced and label propagation, and modularity optimization. We report \textit{VOI} between the natural communities and those identified by the algorithms (results are averages over $1000$ runs).}
\begin{tabular}{cccccc}
\hline\noalign{\smallskip}
\multirow{2}{*}{Network} & \multirow{2}{*}{Number} & \multicolumn{4}{c}{\textit{$VOI$}} \\
 & & \lpa & \bpa & \bpal  & \gmo \\
\noalign{\smallskip}\hline\noalign{\smallskip}
\textit{karate} & $2$ & $0.239$ & $0.145$ & $\mathbf{0.142}$ & $0.218$ \\
\textit{dolphins} & $2$ & $0.363$ & $\mathbf{0.063}$ & $\mathbf{0.062}$ & $0.257$ \\
\textit{football} & $12$ & $\mathbf{0.155}$ & $0.169$ & $0.168$ & $0.323$ \\
\noalign{\smallskip}\hline
\end{tabular}
\end{table}

Note that, in the case of \textit{karate} and \textit{dolphins} networks, balanced propagation performs significantly better than label propagation (and modularity optimization), when in the case of \textit{football} network, the obtained \textit{VOI} is roughly the same. Thus, despite relatively similar performance on synthetic benchmark networks (\secref{expr_sn}), balanced propagation more accurately identifies the true communities within these real-world networks than label propagation (and also modularity optimization). 

For a better comprehension, the \textit{fraction of correctly classified}~\cite{GN02} nodes for \bpal~algorithm equals $72\%$, $96\%$ and $81\%$ for \textit{karate}, \textit{dolphins} and \textit{football} networks respectively (on average).

In~\tblref{rws_qual} we also report average conductance $\overline{\Phi}$ and modularity $Q$ of the revealed community structures for all networks in~\tblref{rws_desc} (mainly to enable comparison with earlier work). Balanced propagation also performs better in terms of conductance. Still, results should be taken with caution as \bpa~and \bpal~algorithms commonly return larger communities than \lpa~algorithm, which implies lower average conductance (see below). On the other hand, according to modularity, performance depends on the size of the network. We argue that this is an artifact of an intrinsic scale incorporated into the measure of modularity (i.e., \textit{resolution limit}~\cite{FB07,GMC10}), thus, lower values of modularity obtained by balanced propagation on smaller networks should not be attributed to weaker community structure (see~\tblref{rws_comm}).

Again, a general pattern can be observed between both balanced propagation algorithms. 

\begin{table}[h]
\centering
\caption{\label{tbl_rws_qual}Analysis of community detection significance of balanced and label propagation. We report the  average conductance $\overline{\Phi}$ and modularity $Q$ of communities identified by different algorithms (results are averages over $1000$ runs).}
\begin{tabular}{cccccccc}
\hline\noalign{\smallskip}
\multirow{2}{*}{Net.} & \multicolumn{3}{c}{\textit{$\overline{\Phi}$}} & \multicolumn{3}{c}{$Q$} \\
 & \lpa & \bpa & \bpal & \lpa & \bpa & \bpal \\
\noalign{\smallskip}\hline\noalign{\smallskip}
\textit{kara.} & $0.285$ & $0.254$ & $\mathbf{0.242}$ & $\mathbf{0.355}$ & $0.296$ & $0.301$ \\
\textit{dolph.} & $0.345$ & $0.082$ & $\mathbf{0.078}$ & $\mathbf{0.485}$ & $0.377$ & $0.380$ \\
\textit{books} & $0.272$ & $\mathbf{0.063}$ & $\mathbf{0.062}$ & $\mathbf{0.505}$ & $0.460$ & $0.460$ \\
\textit{foot.} & $0.328$ & $\mathbf{0.295}$ & $\mathbf{0.296}$ & $0.593$ & $\mathbf{0.602}$ & $\mathbf{0.602}$ \\
\textit{jazz} & $0.210$ & $\mathbf{0.141}$ & $\mathbf{0.142}$ & $\mathbf{0.340}$ & $0.285$ & $0.285$ \\
\textit{eleg.} & $0.354$ & $0.120$ & $\mathbf{0.117}$ & $\mathbf{0.117}$ & $0.036$ & $0.037$ \\
\textit{netsci} & $0.063$ & $\mathbf{0.006}$ & $\mathbf{0.007}$ & $0.879$ & $\mathbf{0.945}$ & $\mathbf{0.944}$ \\
\textit{power} & $0.431$ & $\mathbf{0.129}$ & $\mathbf{0.129}$ & $0.595$ & $\mathbf{0.888}$ & $\mathbf{0.887}$ \\
\noalign{\smallskip}\hline
\end{tabular}
\end{table}

\begin{figure}[t]
\centering
\includegraphics[width=1.00\columnwidth]{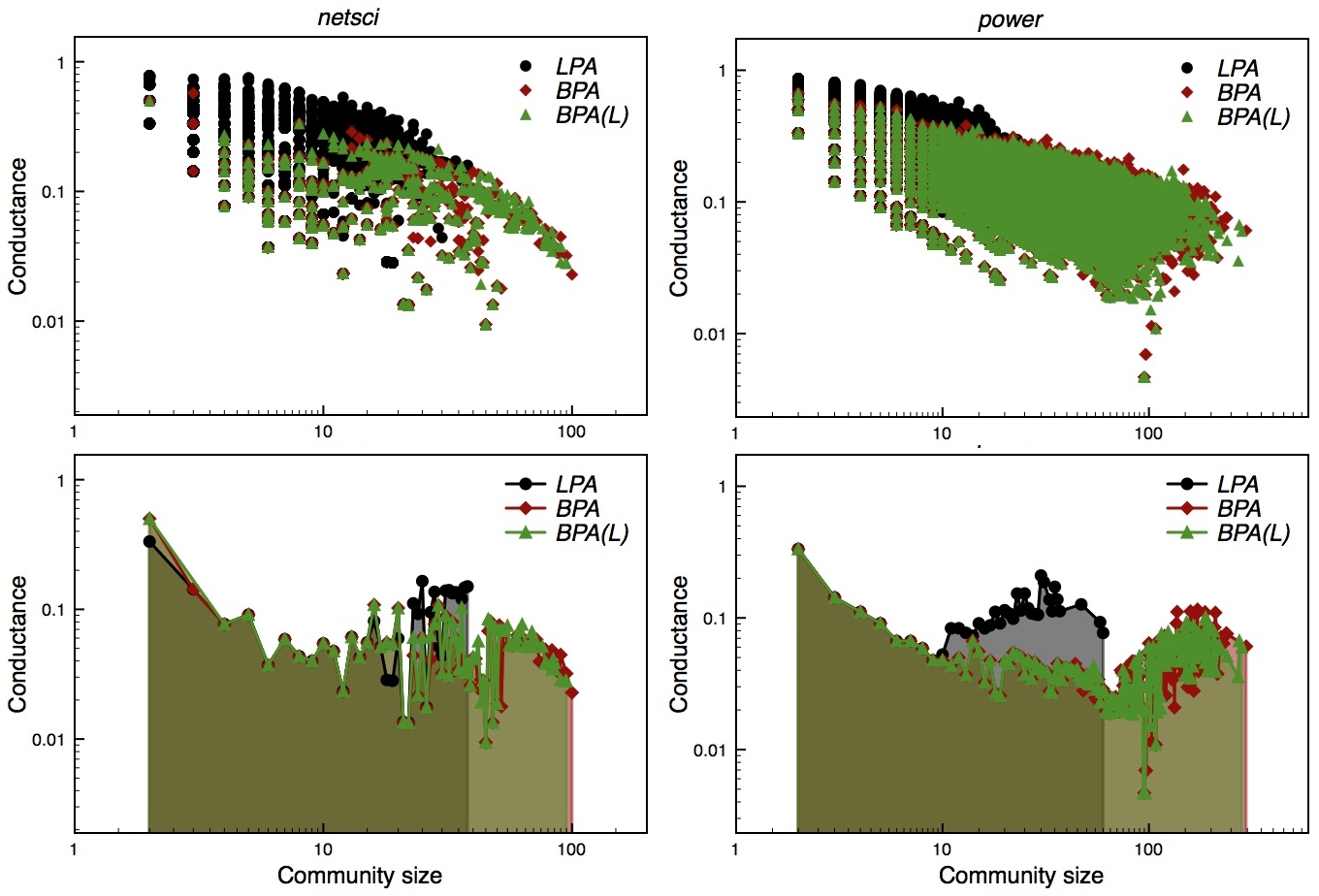}
\caption{\label{fig_rws}(Color online) Comparison of balanced and label propagation on \textit{netsci} and \textit{power} networks. We report the scatter plots showing individual communities, and the minimum values (i.e., lower hulls) at different scales (top, bottom respectively). Results were obtained over $100$ runs.}
\end{figure}

Next, we further analyze the larger two networks in~\tblref{rws_desc}, namely, \textit{netsci} and \textit{power}. We apply each algorithm $100$ times and analyze the conductance of obtained communities at different scales. The results are reported in the form of \textit{network community profile} (\textit{NCP})~\cite{LLDM09} plots, and are shown in~\figref{rws}. \textit{NCP} plots measure the quality of the best community (due to conductance) as a function of its size (\figref{rws}, below). Social and information, and also technological, networks commonly reveal rather characteristic structure of \textit{NCP} plots, with initial decreasing and subsequent increasing trend (for more see~\cite{LLDM09}).

Observe that balanced propagation identifies communities on a much wider scale, including also larger communities. The structure of \textit{NCP} plots thus better coincides with the analysis of~\citet{LLDM09}{Leskovec\etal}, where a natural (i.e., best) community size was estimated to a round $100$ nodes. In other words, basic label propagation finds best communities at much smaller scale than balanced propagation (i.e., at a round $10$ nodes), when the conductance is also significantly higher on average (\tblref{rws_qual}). Note also that label propagation reveals a number of communities with very high conductance (i.e., (black) circles in the uppermost part of~\figref{rws}, top), which can be directly related to the issues of the algorithm discussed in~\secref{lp}.

We conclude that, at least for the networks analyzed, balanced propagation is indeed more stable than basic label propagation, when the quality of the identified community structure is also improved in most cases.

Last, we also briefly analyze the scalability of the proposed balanced propagation. In~\tblref{rws_scal} we report the average number of iterations\footnotemark[2] made by the algorithms over $1000$ runs. As discussed in~\secref{bp}, we do not directly address the issues with overlapping communities. Therefore, nodes, having strong connections with different communities, can prevent basic balanced propagation from converging. The results in~\tblref{rws_scal} thus include only the runs where the algorithms converged in a fixed (maximal) number of iterations (this includes at least $90\%$ of runs in each case). For the same reason, \textit{netsci} and \textit{power} networks were not included in the analysis.

\footnotetext[2]{Each iteration has linear time complexity $\mathcal{O}(|E|)$.}

\begin{table}[h]
\centering
\caption{\label{tbl_rws_scal}Analysis of complexity of balanced and label propagation. We report the average number of iterations made by the algorithms over $1000$ runs (see text).}
\begin{tabular}{cccc}
\hline\noalign{\smallskip}
\multirow{2}{*}{Network} & \multicolumn{3}{c}{Iterations} \\
 & \lpa & \bpa & \bpal \\
\noalign{\smallskip}\hline\noalign{\smallskip}
\textit{karate} & $\mathbf{3.8}$ & $12.6$ & $12.8$ \\
\textit{dolphins} & $\mathbf{4.9}$ & $21.5$ & $22.3$ \\
\textit{books} & $\mathbf{4.9}$ & $31.0$ & $28.8$ \\
\textit{football} & $\mathbf{3.7}$ & $23.4$ & $22.7$ \\
\textit{jazz} & $\mathbf{4.8}$ & $25.9$ & $25.0$ \\
\textit{elegans} & $\mathbf{7.1}$ & $16.1$ & $16.1$ \\
\noalign{\smallskip}\hline
\end{tabular}
\end{table}

The complexity of label propagation is quite lower compared to balanced propagation. Still, all algorithms reveal communities in a relatively small number of iterations and can be easily scaled to larger networks (exhibit near linear time complexity $\mathcal{O}(|E|)$). It should also be noted that extremely fast convergence of label propagation can be somewhat related to random node updates (\secref{lp}). Random update order can be seen as increasing propagation strength from certain nodes (\secref{bp}), which limits the dynamics of the algorithm, and instantly leads it towards some stable, probably suboptimal (i.e., random), partition. The convergence of the algorithm is thus indeed fast, still, the identified community structure is extremely unstable and often suboptimal (as also observed by previous work~\cite{RAK07,TK08,LM09b,SB11}).


\begin{figure*}[t]
\centering
\includegraphics[width=0.85\textwidth]{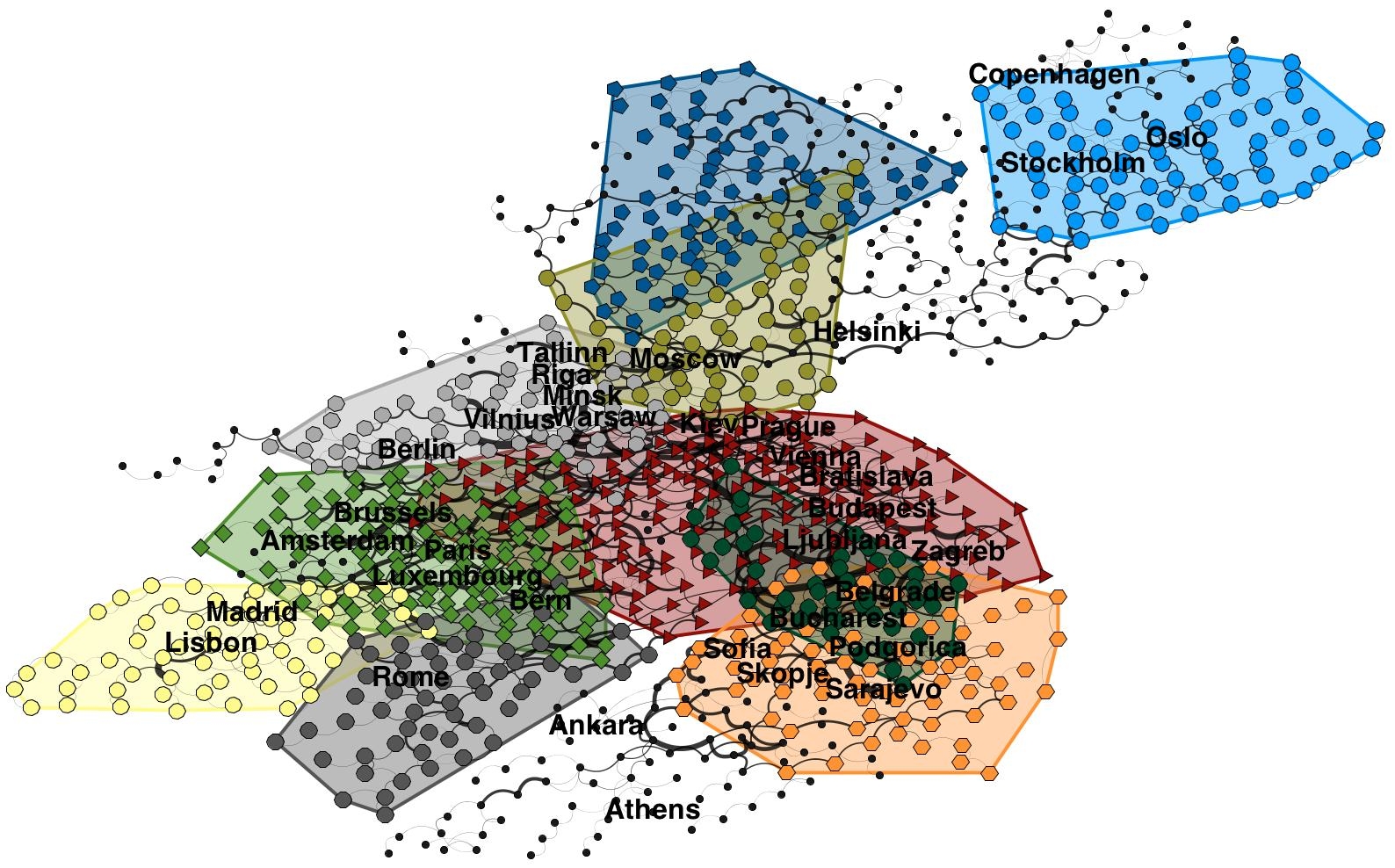}
\caption{\label{fig_er}(Color online) Community structure of the main component of European road network revealed with balanced propagation (i.e., \bpa~algorithm). Node symbols (colors) correspond to different communities, when edge widths represent significant inter-community edges. Due to clarity, only the largest $10$ communities of total $24$ are shown ($Q=0.8344$ and $\overline{\Phi}=0.0796$). Note how communities quite accurately coincide with different (geographical) regions of Europe.}
\end{figure*}

\subsection{\label{sec_expr_ern}European road network}
Road networks are not considered to convey a clear community structure, consisting of densely connected modules (due to sparsity of such networks). However, the network can still contain groups of nodes that are well isolated from others (i.e., connected through only few edges) and community detection algorithms can be employed to reveal such partition of the network. Communities should in this case largely relate to the properties of the road transport within the region, and also coincide with the geographical characteristics of the area.


We have constructed a network of all roads included in the \textit{International E-road Network} (\figref{er}). Nodes thus correspond to European cities and edges represent direct (class A, B) road connections among them. We limit the analysis to the main component of the network that consists of $1039$ nodes and $1355$ edges (a complete network has $1177$ nodes and $1469$ edges). Note that the network is neither \textit{scale-free}~\cite{BA99} (i.e., maximum degree equals $10$, when the degree distribution is, e.g., log-normal) nor \textit{small-world}~\cite{WS98} (i.e., average distance among nodes is $l=18.40$ and the clustering coefficient~\cite{WS98} equals $C=0.02$).

Due to long average distances among different parts of the network, road networks are particularly hard to partition with standard community detection algorithms. Furthermore, as the network has almost tree-like structure, it is often hard to decide where to split long paths of nodes. Indeed, if we apply the basic label propagation (i.e., \lpa~algorithm) we obtain $343$ communities with $Q=0.5617$ and $\overline{\Phi}=0.4424$ (on average over $1000$ runs). Hence, communities consist of only $3.03$ nodes on average, thus, they can only hardly be considered as meaningful.

On the other hand, balanced propagation (i.e., \bpa~algorithm) partitions the network into $35$ communities with $Q=0.8374$ and $\overline{\Phi}=0.1224$ (on average over $1000$ runs). In~\figref{er} we show the community structure that obtained minimum average conductance $\overline{\Phi}$. Note how the largest communities quite accurately coincide with different (geographical) regions of Europe. In particular, from left to right (top to bottom), communities represent cities of Iberian Peninsula (e.g., Madrid), eastern Central Europe (e.g., Berlin), western Central Europe (e.g., Paris), Apennine Peninsula (e.g., Rome), eastern Russia, western Russia and Finland (e.g., Moscow), northern East Europe (e.g., Bratislava), southern East Europe (e.g., Bucharest), Balkan Peninsula (e.g., Skopje), Scandinavian Peninsula (e.g., Stockholm), etc. It is ought to be mentioned that, although community structures revealed by the algorithm through different runs indeed differ, in most cases, largest communities correspond to the same regions as discussed above. The latter thus further confirms the robustness of the balanced propagation.



\section{\label{sec_conc}Conclusions}
The article addresses one of the main issues of label propagation algorithm for community detection -- the stability of the identified community structure. We introduce balanced propagation that controls  (i.e., stabilizes) the dynamics of basic label propagation through utilization of node balancers. The resulting approach is significantly more robust than its label propagation counterpart, when its community detection strength is even improved. Thus, balanced propagation retains high scalability and algorithmic simplicity of label propagation, but improves on its stability and performance. The proposition has been validated on synthetic networks with planted partition, and on several real-world networks with community structure. Moreover, the proposed algorithm was further applied to an entire European road network, where it accurately partitions the network with respect to (geographical) regions.

Due to its simplicity, balanced propagation can be easily incorporated into arbitrary (label) propagation algorithm, not limited to the field of community detection. Moreover, the work provides further comprehension of the propagation on networks, with different applications.


\begin{acknowledgement}
The work has been supported by the Slovene Research Agency \textit{ARRS} within the research program P2-0359.
\end{acknowledgement}


\bibliographystyle{epj}


\end{document}